\documentclass
[preprint,pre,groupedaddress,nobibnotes,nofootinbib,notitlepage,10pt,twocolumn,showpacs]{revtex4}%
\usepackage{amsfonts}
\usepackage{amsmath}
\usepackage{amssymb}
\usepackage{graphicx}%
\begin{document}
\title{On the dynamical anomalies \\in numerical simulations of selfgravitating systems.}
\author{L. Velazquez}
\email{luisberis@geo.upr.edu.cu}
\affiliation{Departamento de F\'{\i}sica, Universidad de Pinar del R\'{\i}o, Mart\'{\i}
270, Esq. 27 de Noviembre, Pinar del R\'{\i}o, Cuba.}
\author{F. Guzm\'{a}n}
\email{guzman@info.isctn.edu.cu}
\affiliation{Departamento de F\'{\i}sica Nuclear, Instituto Superior de Ciencias y
Tecnolog\'{\i}a Nucleares, Carlos III y Luaces, Plaza, La Habana, Cuba.}
\date{\today}

\begin{abstract}
According to self-similarity hypothesis, the thermodynamic limit could be
defined from the scaling laws for the system self-similarity by using the
microcanonical ensemble. This analysis for selfgravitating systems yields the
following thermodynamic limit: $N\rightarrow\infty$ keeping constant
$E/N^{\frac{7}{3}}$ and $LN^{\frac{1}{3}}$, in which is ensured the
extensivity of the Boltzmann entropy $S_{B}=\ln W\left(  E,N\right)  $. It is
shown how the consideration of this thermodynamic limit allows us to explain
the origin of dynamical anomalies in numerical simulations of selfgravitating systems.

\end{abstract}
\maketitle

\section{Introduction}

In the last years there is a special interest to perform a well-defined
thermodynamical description for selfgravitating systems and many interesting
results have been obtained
\cite{pad,kies,ruelle,lind,antonov,Lynden,Lynden1,thirring,chava,gross}. In
ref.\cite{sanch}, de Vega and Sanchez have pointed out that a kind of
thermodynamical limit of a self-gravitating system can be defined if one
considers what they call the diluted limit: send the number of particles, $N$,
and the volume, $V$ , to infinity, keeping constant the ratio $N/V^{1/3}$
instead of the density, $N/V$. Recently, V. Laliena in ref.\cite{laliena}
showed that this thermodynamic limit suffers from the same problems as the
usual thermodynamical limit and leads to divergent thermodynamical functions.

In the recent paper \cite{vel6}, we proposed a new thermodynamic limit for
selfgravitating system which is obtained from the \textit{self-similarity
properties} \cite{vel1,vel2,vel3,vel4,vel5} of this kind of systems:%

\begin{equation}
N\rightarrow\infty\text{, keeping constant }\frac{E}{N^{\frac{7}{3}}}\text{
and }LN^{\frac{1}{3}}. \label{thermo limit}%
\end{equation}
As already shown in ref.\cite{vel6}, the consideration of this thermodynamic
limit allows us to perform a well-defined thermodynamical description for
selfgravitating systems by using almost the same arguments used by the
standard Thermodynamics for describing the extensive systems. The aim of this
paper is to show that the dynamical anomalies in numerical simulations of
selfgravitating systems disappear when the self-similarity properties of this
kind of systems are taken into account.

\section{What is the system self-similarity?}

Self-similarity is a natural generalization of the extensivity of the
traditional system, which should be taken into account in order to perform a
well-defined thermodynamical description based on the microcanonical ensemble
\cite{vel1,vel2,vel3}. Basically, it can be obtained from the asymptotical
scaling behavior of thermodynamical variables and the microcanonical
phase-space volume $W$ when the many particle limit $N\rightarrow\infty$ is invoked.

There are many studies in long-range Hamiltonian systems where the validity of
the thermodynamical description in the thermodynamic limit is intimately
related to the scaling with $N$ of the thermodynamic variables and potentials.
The problems with the nonextensive nature of those systems are avoided using
the physically unjustified Kac prescription \cite{kac} in which the coupling
constants are scaled by some power of $N$ in order to deal with an extensive
total energy $E$. Self-similarity can not be reduced to the Kac prescription,
because the energy extensivity is not demanded in the self-similarity
framework. Geometrical aspects of the probabilistic distribution function of
the statistical ensembles allows us to define a well-defined thermodynamic
formalism even in cases where the extensive nature of systems can not be
ensured \cite{vel1,vel2,vel6}.

The type of self-similarity scaling laws determines which is the generalized
Boltzmann entropy \cite{vel1}\ which should be considered in the application
in order to guarantee the ensemble equivalence of the microcanonical ensemble
with some adequate generalization of the canonical ensemble \cite{vel1}.
Ordinarily, scaling laws are \textit{exponential}, it means that the
microcanonical accessible volume $W$ has an exponential growing with $N$ in
the thermodynamic limit. There is not \textit{a priory} reason for supposing
that a Hamiltonian system should necessarily satisfy this kind of growing in
the thermodynamic limit, although there is a huge world of Hamiltonian systems
that do obey it. In ref.\cite{vel3} we used the self-similarity hypothesis in
order to find the necessary conditions for the validity of Tsallis' Statistics
\cite{tsal} by considering these arguments and using a procedure inspired in
the one used by Gross in deriving his microcanonical thermostatistical theory
\cite{gross2}. It was proved in that paper the nonuniqueness of a
thermodynamic formalism based on the microcanonical ensemble. In this sense,
self-similarity also difers from the Kac prescription.

\section{The self-similarity scaling laws}

In this section is recalled how can be obtained the thermodynamic limit
(\ref{thermo limit}) by considering the system self-similarity properties for
the following tridimensional selfgravitating Hamiltonian system:%

\begin{equation}
H=T+U=\underset{k=1}{\overset{N}{\sum}}\frac{1}{2m}\mathbf{p}_{k}%
^{2}+\underset{j>k=1}{\overset{N}{\sum}}\phi\left(  r_{jk}\right)  .
\label{hamilton}%
\end{equation}
Here $r_{jk}=\left\vert \mathbf{r}_{j}-\mathbf{r}_{k}\right\vert $ is the
distance between the \textit{j-th} and \textit{k-th} particles, being
$\phi\left(  r\right)  $ the selfgravitating potential which behaves as
Newtonian interaction when $r\rightarrow\infty$:%

\begin{equation}
\phi\left(  r\right)  \sim1/r. \label{beh r}%
\end{equation}
It is easy to show that the microcanonical phase-space accessible volume $W$
is given by:%

\begin{equation}
W=\frac{1}{N!\Gamma\left(  \frac{3N}{2}\right)  }\int_{V^{N}}\underset
{k=1}{\overset{N}{\prod}}d^{3}\mathbf{r}_{k}\left[  E-U\right]  ^{\frac{3N}%
{2}-1},
\end{equation}
where $V$ is the tridimensional volume in which the system is enclosed. Since
$\phi\left(  r\right)  $ describes a long-range interaction, the potential
energy $U$ for $N$ large could be estimated as follows:%

\begin{equation}
U\sim N^{2}L^{-1},
\end{equation}
where $L$ is the characteristic linear dimension of the system. The factor
$N^{2}$ takes into account the contribution of all couples of particles, while
the factor $L^{-1}$ appears as consequence of the behavior of the potential
$\phi\left(  r\right)  $ (\ref{beh r}) for $r$ large . The total energy $E$
will exhibit an identical behavior:%

\begin{equation}
E\sim N^{2}L^{-1}.
\end{equation}
Thus, when $N$ is sent to infinity, the microcanonical phase-space accessible
volume $W$ depends on $N$ and $L$ as follows:%

\begin{equation}
W\sim\frac{1}{N!\Gamma\left(  \frac{3N}{2}\right)  }\left[  N^{2}%
L^{-1}\right]  ^{\frac{3N}{2}}L^{3N}\sim\left[  LN^{\frac{1}{3}}\right]
^{\frac{3N}{2}}.
\end{equation}
Thus, when the thermodynamic limit (\ref{thermo limit}) is taken into account,
the Boltzmann entropy is \textit{extensive}:%

\begin{equation}
S_{B}=\ln W\sim N.
\end{equation}

There are some connections suggesting the validity of the thermodynamic limit
(\ref{thermo limit}).  It was proposed in ref.\cite{vel5}, an alternative
model for the classic isothermal model of Antonov \cite{antonov} which uses an
energetic prescription instead of a box renormalization in order to avoid the
long-range singularity of the \textit{N}-body selfgravitating systems. In that
paper, a classical gas of identical particles with mass $m$ was considered.
Taking into account the rest energy of the particles, the nonrelativistic
limit is valid when the absolute value of the mechanical energy of the system
is much smaller than its rest energy :%

\begin{equation}
\left\vert \epsilon_{0}\epsilon\left(  \Phi_{0}\right)  N^{\frac{7}{3}%
}\right\vert \ll mc^{2}N,
\end{equation}
where $\epsilon_{0}$ is the characteristic energy of this model:%

\[
\epsilon_{0}=2\pi\frac{G^{2}m^{5}}{\hbar^{2}}.
\]
However, this condition can not be satisfied for an arbitrary number of
particles. In fact, when $N$\ tends to $N_{0}$:%

\begin{equation}
N_{0}=\left(  2\pi\frac{mc^{2}}{\epsilon_{0}}\right)  ^{\frac{3}{4}}=\left(
\frac{\hbar c}{G}\right)  ^{\frac{3}{2}}\frac{1}{m^{3}},
\end{equation}
which corresponds to a characteristic mass $M_{0}$:%

\begin{equation}
M_{0}=N_{0}m=\left(  \frac{\hbar c}{G}\right)  ^{\frac{3}{2}}\frac{1}{m^{2}},
\end{equation}
the model loses its validity. Everybody can recognize the fundamental constant
of the stellar systems, which has much to do with the stability conditions of
the stars (see in ref.\cite{chand}). Note that this constant appears as a
consequence of assuming the energy \textit{N}-dependence (\ref{thermo limit}),
so that, it could not have been obtained if another thermodynamic limit had
been adopted. A consequent analysis of these massive systems should be
performed taking into account the relativistic effects.

Another link is found in the well-known white dwarfs model based on the
consideration of the Thomas-Fermi method to describe the state equation of the
degenerate nonrelativistic electronic gas, whose pressure supports the
hydrostatic equilibrium of the star:%

\begin{equation}
\Delta\phi=-4\pi G\rho,
\end{equation}
where $\phi$ is the Newtonian potential, being $\rho$ the mass density:%

\begin{equation}
\rho=\mu m\frac{2^{\frac{3}{2}}m_{e}^{3}\left(  \phi_{S}-\phi\right)
^{\frac{3}{2}}}{3\pi^{2}\hbar^{3}},
\end{equation}
where $\mu$ is the number of nucleons per electron, $m$ and $m_{e}$ are the
nucleon and electron masses, being $\phi_{S}$ the potential at the star
surface. From this model it can be easily derived the \textit{N-}dependence of
total energy and linear dimension of the system given in (\ref{thermo limit})
by using a simple dimensional analysis. This coincidence is not casual: for
the nonrelativistic particles the thermodynamic limit only depends on the
dimension of the physical space.

According to the exposed above, during the many particle limit, $N\rightarrow
\infty$, the selfgravitating Hamiltonian systems (\ref{hamilton}) exhibit
\textit{self-similarity} under the following scaling laws:%

\begin{equation}
\left.
\begin{array}
[c]{c}%
N_{o}\rightarrow N\left(  \alpha\right)  =\alpha N_{o,}\\
E_{o}\rightarrow E\left(  \alpha\right)  =\alpha^{\frac{7}{3}}E_{o},\\
L_{o}\rightarrow L\left(  \alpha\right)  =\alpha^{-\frac{1}{3}}L_{o},
\end{array}
\right\}  \Rightarrow W\left(  1\right)  \rightarrow W\left(  \alpha\right)
=\mathcal{F}\left[  W\left(  1\right)  ,\alpha\right]  , \label{scaling laws}%
\end{equation}
where the functional $\mathcal{F}\left[  W,\alpha\right]  $ defines an
\textit{exponential scaling laws}:%

\begin{equation}
\mathcal{F}\left[  W,\alpha\right]  =\exp\left[  \alpha\ln W\right]  ,
\label{exponential ssl}%
\end{equation}
being
\begin{equation}
W\left(  \alpha\right)  =W\left[  E\left(  \alpha\right)  ,N\left(
\alpha\right)  ,L\left(  \alpha\right)  \right]  .
\end{equation}

As already mentioned in our previous works \cite{vel1,vel2,vel3,vel4}, a
system exhibits self-similarity under certain scaling transformations of its
fundamental macroscopic observables when the functional $\mathcal{F}\left[
W,\alpha\right]  $ obeys to the general condition of self-similarity
\cite{vel1}:%

\begin{equation}
\mathcal{F}\left[  \mathcal{F}\left[  x,\alpha_{1}\right]  ,\alpha_{2}\right]
=\mathcal{F}\left[  x,\alpha_{1}\alpha_{2}\right]  .
\end{equation}
Only in this case, the scaling transformations constitute a group of
transformations compatible with the system macroscopic properties, whose
ordering information is contained in the microcanonical accessible phase space
volume $W$. We suppose that when the thermodynamic limit is carried out by
using these special scaling laws:

\begin{itemize}
\item all those system self-similarity properties are protected, even though,
its dynamic characteristics,

\item as well as under certain conditions, the microcanonical ensemble could
be \textit{equivalent} to certain generalization of canonical ensemble.
\end{itemize}

Contrary, we guess that the non consideration of the self-similarity scaling
laws should provoke a \textit{trivial ensemble inequivalence}, as well as
others anomalous behaviors \cite{vel4}. According to our viewpoint, the
Boltzmann's definition of entropy is applicable to this system, which allows
us to classify the selfgravitating model as a pseudoextensive system
\cite{vel2}.

\section{Discussions and conclusions}

Many investigators do not pay so much attention to the analysis of the system
self-similarity when it is performed its macroscopic description. However, the
non or bad consideration of the system self-similarity scaling laws leads in
many cases to the trivial \textit{nonequivalence of the statistics ensembles}
as well as other anomalous behaviors, such as the reduction of the mixing and
slow relaxation regime towards the equilibrium.

As an example of the above affirmation are the anomalies presented in the
dynamical study of the selfgravitating systems performed by Cerruti-Sola \&
Pettini in the ref.\cite{pet}. In that paper, they observed a
\textit{weakening} of the system chaotic behavior with the increasing of the
particles number $N$. It is very well-known the consequence of this fact on
the ergodicity of the system: the chaotic dynamics provides the mixing
property in the phase space necessary for obtaining the equilibrium. In that
example the chaoticity time grows with $N$, and therefore, the systems could
expend so much time to arrive to the equilibrium.

In a recent paper \cite{taruya}, Taruya and Sakagami carried out a $N$-body
numerical experiment of the same situation as investigated in classic papers
\cite{antonov,Lynden}. That is, they confined the $N$ particles interacting
via Newton gravity in a spherical adiabatic wall, which reverses the radial
components of the velocity if the particle reaches the wall. The typical
timescales appearing in this system are the free-fall time, $T_{ff}%
=(G\rho)^{-1/2}$, and the global relaxation time driven by the two-body
encounter, $T_{relax}=(0.1N/\ln N)T_{ff}$ \cite{binney}. Since they considered
the density $\rho$ as \textit{N}-independent quantity, the relaxation time
$T_{relax}$ diverges when the thermodynamic limit is invoked.

These facts can be interpreted as the non-commutativity of the thermodynamic
limit ($N\rightarrow\infty)$ with the infinite time limit ( $t\rightarrow
\infty$): when the first is performed before the second, the systems will not
relax to the microcanonical ensemble. However, our analysis allows\ us to
understand the origin of all these anomalous behaviors.

It was considered in the study \cite{pet} that the energy is scaled
proportional to $N$ during the realization of the thermodynamic limit, which
does not take into account the system self-similarity. It is very easy to see
that the instability exponent $\lambda_{H}$ \textit{versus} energy dependence
(Figure 4 of the ref.\cite{pet})\textit{ is corrected} when the
self-similarity scaling law for the energy is considered: its proportionality
to $N^{\frac{7}{3}}$. The weakening of chaoticity\textit{ disappears }when
self-similarity is taken into account. Anyway, in spite of they assumed a
different scaling law for the energy, they obtained the correct dependency of
the instability exponent $\lambda_{H}$ with the energy per particle: the power
law $\epsilon^{\frac{3}{2}}$.

The dynamical anomalies in the Taruya and Sakagami analysis are also explained
by the non consideration of self-similarity. When the thermodynamic limit
(\ref{thermo limit}) is taken into consideration, system volume scales as
$\sim1/N$, and therefore, the characteristic density $\rho$ scales as $\sim
N^{2}$. Thus, $T_{ff}\sim1/N$, and hence, $T_{relax}\sim1/\ln N\rightarrow0$.
So that, the origin of the dynamical anomalies is a trivial consequence of the
consideration of a thermodynamic limit incompatible with the system
self-similarity properties of the self-gravitating systems.

There are some other examples of dynamical anomalies which at first glance
could lead to the inapplicability of the microcanonical or the canonical
ensemble (see for example in
refs.\cite{lat1,lat4,lat2,lat3,zanette,campa1,campa2,kandru}). However, it
could be proved that all these dynamical anomalies can be explained by the
same arguments exposed in this paper.

\end{document}